\documentclass[11pt]{book}
\usepackage{Wiley-AuthoringTemplate}
\usepackage{chapterbib}
\usepackage[sectionbib,authoryear]{natbib}
\usepackage{fullpage}

\newcommand{\link}{{\mathcal L}}
\newcommand{\flh}{{\mathcal A}}
\newcommand{\avec}{\vec A}
\newcommand{\xtwo}{{\mathsf x}}

\renewcommand\div{{\bf \nabla} \cdot }

\newcommand\cross{\times}
\newcommand\grad{ {\bf \nabla } }
\newcommand\curl{ {\bf \nabla} \times}

\newcommand\xhat{\hat x}
\newcommand\yhat{\hat y}
\newcommand\zhat{\hat z}
\newcommand\rhat{\hat r}
\newcommand\phihat{\hat \phi}
\newcommand\thetahat{\hat \theta}
\newcommand\nhat{{\hat n}}
\newcommand\oerpi{ \frac{1 }{ 2 \pi } }
\newcommand\half{ \frac{1}{2} }

\newcommand\dotn{\cdot \hat n |_S}

\newcommand\rmd{ {\mathrm d} }
\newcommand\pder[2]{ \frac{\partial #1}{\partial #2} }

\newcommand\deriv[2]{ \frac{\rmd #1}{\rmd #2} }

\newcommand\vol{{\cal V}}

\newcommand\dv{~{\mathrm d}^3 x}
\newcommand\da{~{\mathrm d}^2 x}

\newcommand{\dl}{~{\mathrm d}l}
\newcommand{\dline}{\mathrm{d} \mathbf{l}}

\newcounter{problemnumber}

\newcommand\wind{\hbox{\textit w}}

\newcommand\invcurl{{\rm curl}^{-1}}
\def\da{~d^2 x}

\newcommand\ap{{\bf A_P}}

\makeindex

\begin{document}

\mainmatter
\chapter[Introduction to Field Line Helicity]{Introduction to Field Line Helicity}

\author*[1]{A. R. Yeates}
\author[2]{M. A. Berger}

\address[1]{\orgdiv{Department of Mathematical Sciences}, 
\orgname{Durham University}, \street{South Road}, \city{Durham},
\postcode{DH1 3LE}, \country{UK}}%

\address[2]{\orgdiv{Department of Mathematics}, 
\orgname{University of Exeter}, 
\postcode{EX4 4QF}, 
     \city{Exeter}, \street{North Park Road}, \country{UK}}%

\address*{Corresponding Author: \email{anthony.yeates@durham.ac.uk}}
     
\maketitle

\begin{abstract}{}

Field line helicity measures the net linking of magnetic flux with a single magnetic field line.
It offers a finer topological description than the usual global magnetic helicity integral, while still being invariant in an ideal evolution unless there is a flux of helicity through the domain boundary.
In this chapter, we explore how to appropriately define field line helicity in different volumes in a way that preserves a meaningful topological interpretation.
We also review the time evolution of field line helicity under both boundary motions and magnetic reconnection.
\end{abstract}

\keywords{sun, helicity}

\section{Definitions of field line helicity} \label{sec:defs}

We briefly review topological measures of magnetic field structure. Some of these refer to the structure of the total field within a volume (the \emph{magnetic helicity}), others to the relationship between individual pairs of field lines (linking and winding). Field line helicity is intermediate between these ideas, as it measures the net linking or winding of one field line with the total field. Helicity integrals depend on both the tangling of field lines with each other, as well as the topology and geometry of the volume in which the field lines reside. A strong rationale for considering topological measures is that field line topology is conserved in ideal evolution of the field \citep{Moffatt69}. 

\subsection{Definitions of field line helicity for closed volumes}
\begin{figure}[t]
\begin{center}
\includegraphics[scale=0.3]{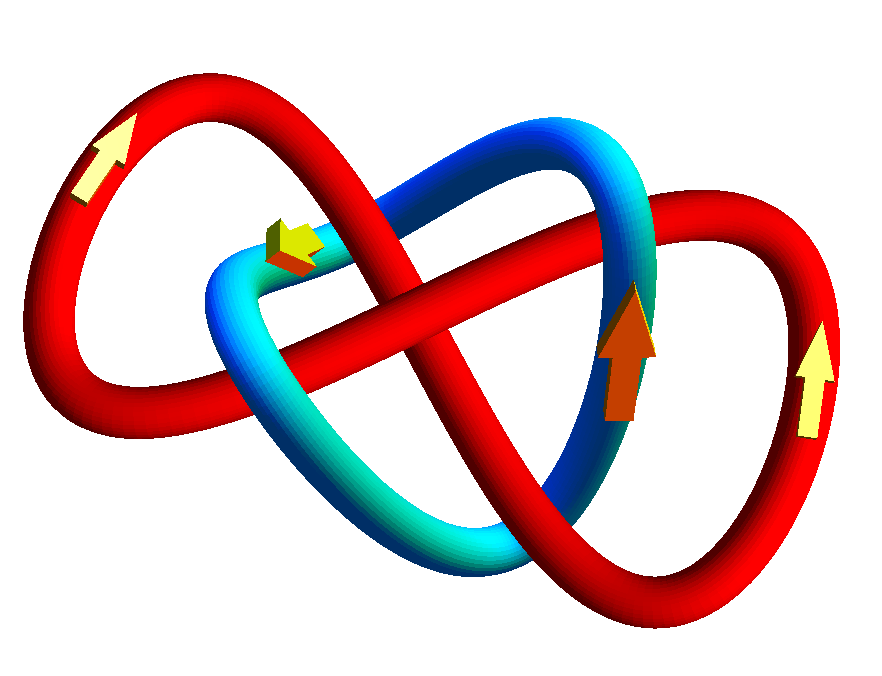}
\end{center}
\caption{For this link (the Whitehead link) the Gauss linking number, $\link_{12}$, is zero. In general, the linking number equals half the difference between the number of positive crossings and the number of negative crossings, as seen in any plane projection.}
\label{fig:linking} 
\end{figure}

The simplest situation occurs with two closed field lines residing within a volume $\vol$ (or infinite space) where the field lines do not cross the boundary of $\vol$.
Historically, Gauss (1809) discovered a double line integral which measures the linking of two
closed curves $L_1$ and $L_2$. Let positions on the curves be given by $\vec x(\sigma)$ and $\vec y(\tau)$. Then
\begin{equation}
\link_{12}  = -\frac{1}{4 \pi} \oint_{L_2}
\oint_{L_1}\deriv {\vec x} {\sigma} \cdot \frac{\vec r}{r^3} \cross
\deriv{\vec y} {\tau} \ d\sigma\ d\tau, \label{gaussvw}
\end{equation}
where ${\vec r} = {\vec x} - {\vec y}$. This may be calculated by counting signed crossings (see Figure \ref{fig:linking}). 

For a magnetic field consisting of a finite collection of closed magnetic flux tubes (a very special case), we can define an overall invariant
\begin{eqnarray}
H_N = \sum_{i=1}^N\sum_{j=1}^N \link_{ij}\Phi_i\Phi_j,
\label{hn}
\end{eqnarray}
where $\Phi_i$ represents the magnetic flux of each tube and $\link_{ij}$ is the linking number between the tubes. This invariant is called the magnetic helicity. 
For a more general magnetic field that nevertheless consists entirely of closed field lines (still a special case), we can take $N\to\infty$ in equation \eqref{hn} so that the sums become integrals. Accounting for the magnetic flux, the tangent vectors in \eqref{gaussvw} become magnetic field vectors and the magnetic helicity may be written
\begin{equation}
H  = -\frac{1}{4 \pi} \int_\vol
\int_\vol \vec B (\vec x) \cdot \frac{\vec r }{ r^3} \cross
B(\vec y) \, \rmd ^3 x \ \rmd ^3 y . \label{totalh}
\end{equation}
Here the integral is over all space, or over a volume containing all field lines.
In fact, in the above equation we do not need to assume that the field lines actually close upon themselves –- some may ergodically fill a subvolume, or twist around a toroidal surface with an irrational winding number. Such fields can be constructed as the limit as $N\rightarrow \infty$ of a sequence of fields consisting of $N$  thin closed flux tubes \citep{Arnold98}. Thus, whether or not the field lines close upon themselves, we can approximate any magnetic field as a collection of such tubes.

Returning to the finite set of $N$ flux tubes with fluxes $\Phi_i$, notice that equation \eqref{hn} can be written as
\begin{equation}
H_N = \sum_{i=1}^N \flh_i \Phi_i; \qquad \flh_i = \sum_{j=1}^N \link_{ij}\Phi_j.
\end{equation}
The limit as $N\to\infty$ of $\flh_i$ is in itself another topological measure, this time defined for every individual magnetic field line -- this is what we call the \emph{field line helicity}.

Since the field lines do not cross the boundary of the volume, we can now use the Biot-Savart formula 
\begin{equation}
\vec A(\vec x) =  
\invcurl \vec B (\vec x) \equiv - \frac{1}{4 \pi}\int_\vol \frac{\vec r }{r^3} \cross \vec B(\vec y)\ \rmd ^3 y
\end{equation}
to find 
\begin{equation}
H =
\int_\vol \vec A
\cdot \vec B\ \rmd ^3 x. 
\label{hab}
\end{equation}
Historically, invariance of the expression \eqref{hab} was known before the interpretation in terms of Gauss linking number was identified \citep{Moffatt69, Moffatt1992}. For a closed line $L$ within the field, the equivalent expression for the field line helicity will simply be
\begin{equation}
\flh (L) = \oint_L \avec \cdot \dl.
\end{equation}
This is just the net magnetic flux encircled by $L$, and since $L$ is a closed curve, it is manifestly gauge invariant.
We remark that \citet{Yahalom2013} interprets $\mathcal{A}(L)$ as a magnetohydrodynamic analogue of the Aharonov-Bohm effect from quantum mechanics: if $\vec{B}=\nabla\chi\times\nabla\eta$ then $\avec = \chi\nabla\eta + \nabla\zeta$, so that a non-zero $H$ requires jumps in $\zeta$ around closed field lines, such jumps giving $\mathcal{A}(L)$.

Suppose our volume $\vol$ is simply connected (e.g. a sphere, not a torus). 
Note here that because $L$ is closed, we can gauge transform $\avec \rightarrow \avec + \grad \phi$ for an arbitrary gauge function $\phi$ without affecting  $\flh (L)$. If we have a multiply connected volume a difficulty arises: the wrong choice of $\avec$ may imply the existence of magnetic flux in the external region that threads through a hole, which makes $\oint_L \avec \cdot \dl$ dependent on the unknown external field.
To remedy this, one can restrict the gauge of $\avec$ so that $\oint_L \avec \cdot \dl = 0$ for any closed curve on the boundary encircling a hole the long way around \citep[see also][]{2019JPlPh..85e7701M}.

\subsection{Definitions of field line helicity for open volumes} \label{sec:opendefs}

\begin{figure}[t]
\begin{center}
\includegraphics[scale=0.5]{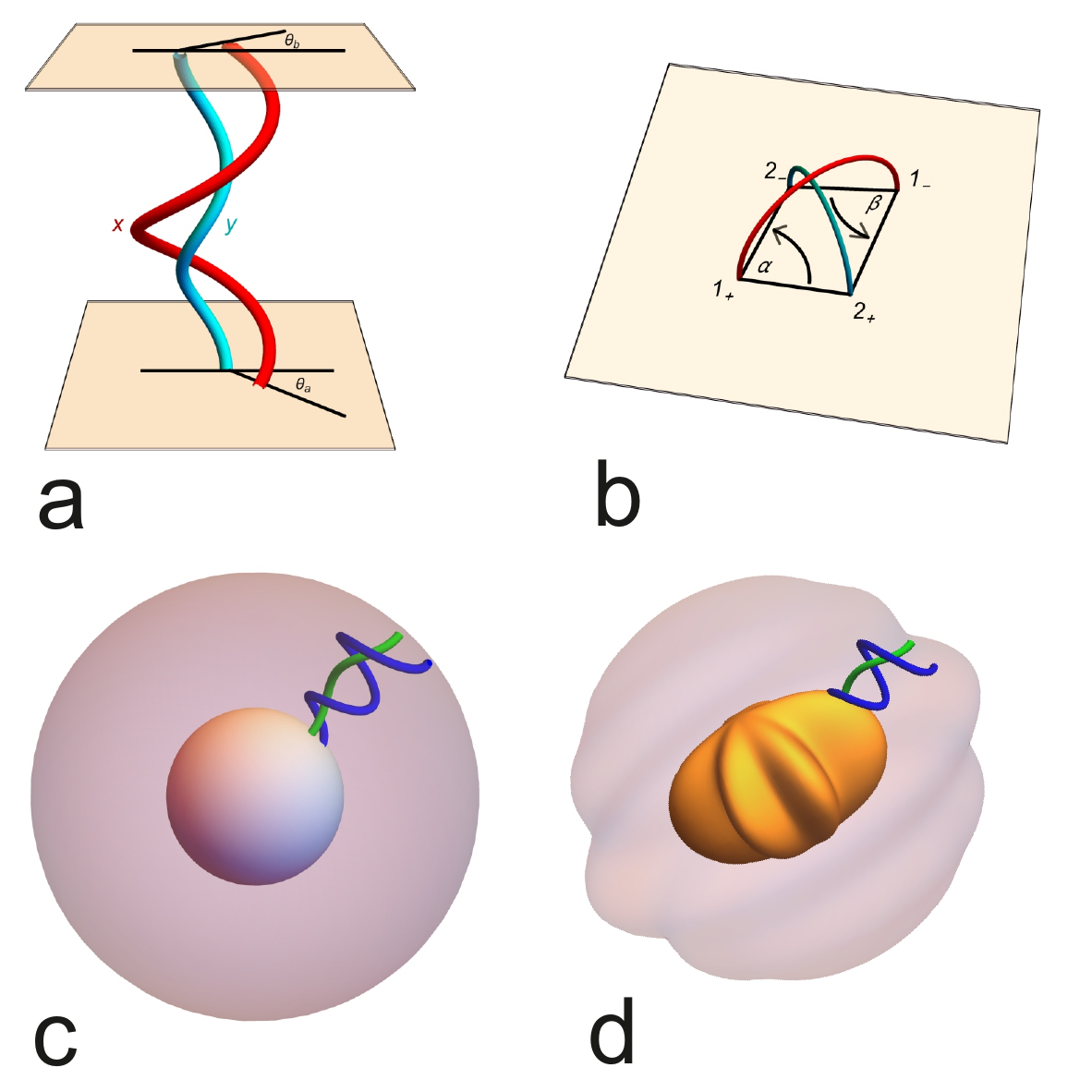}
\end{center}
\caption{Examples of a pair of magnetic field lines in different domains, including (a) the volume between two parallel planes; (b) the volume above a plane; (c) a spherical shell (the volume between two concentric spheres); and (c) the volume between two more general surfaces.}
\label{fig:winding} 
\end{figure}

We now go to volumes with flux crossing the boundary. Suppose a field line forms a loop, with both ends on the same boundary (as for example in Figure \ref{fig:winding}b). An early approach \citep{Antiochos1987} involves drawing a geodesic between the two endpoints, thus forming a closed curve for which  $\oint \avec \cdot \dl$ can be measured. This scheme does not integrate to the full helicity, however. 
We wish to define field line helicity to be consistent with full helicity, but also so that gauge transformations of $\avec$ do not change our results. There are two principle approaches to this problem. One consists of defining topological quantities such as winding number, then summing to find the helicities. If the boundary flux does not move, then ideal motions of the field lines will not change the windings, so field line helicity will be conserved. If the boundary flux does move, then topological structure can cross the boundary leading to changes in helicity measures. This will be discussed in Section \ref{sec:ideal}.

The second approach is to measure helicity \emph{relative} to a potential field, i.e. a field with zero current in the volume. As the potential field is the minimum energy state for given boundary conditions, its structure only depends on the shape of the boundary and the distribution of flux on the boundary (and for multiply connected volumes, specified new fluxes around each hole). For highly symmetric volumes such as those bounded by planes or spheres, the two approaches give identical results.

For more complicated volumes, the shape of the boundary or boundaries can add extra terms which distinguish the two approaches, as will be discussed in Section \ref{sec:gen}. A definition of field line helicity can also be based on relative helicity \citep{Moraitis2019}, to be discussed in Section \ref{sec:rflh}.


\subsubsection{Volumes between parallel planes}

Suppose the volume of interest $\vol$ is the space between parallel planes, say $z=0$ and $z= h$, as illustrated in Figure \ref{fig:winding}a. We will first assume that all field lines start at $z=0$ and end at $z=h$. Then no two field lines link in the Gauss sense; however, they will twist about each other through some angle $\delta \theta_{ij}$. The \emph{winding number} makes this an integer for complete twists,
\begin{equation}
\wind_{ij} = \frac{1}{2\pi} \delta \theta_{ij}.
\end{equation} 
For the example in Figure \ref{fig:winding}a, we have $\wind_{ij} = ( \theta_b-\theta_a )/(2\pi) + 1$.

We can now adopt formulae for the helicity and field line helicity similar to the closed field case, with winding number replacing linking number. We denote horizontal positions by
$\xtwo = (x, y)$. Also, we distinguish positions in the bottom $x-y$ plane by $\tilde {\xtwo} = (x,y,0)$. Let $L_i$ be the field line with foot point at $\tilde{\xtwo}_i$. Also let $\wind(\tilde {\xtwo_i}, \tilde {\xtwo_j})$ be the winding number between $L_i$ and $L_j$. Then
the field line helicity of $L_i$ will be 
\begin{eqnarray}
\flh(L_i) = \int \wind(\tilde \xtwo_i, \tilde \xtwo_j) B_z(\tilde \xtwo_j) \ \rmd ^2 \tilde \xtwo_j.
\end{eqnarray}

Let us build up $\flh(L_i)$ one plane at a time; in other words find expressions for 
$\rmd \flh(L_i)/\rmd z$.
For now we make the simplifying assumption that $B_z>0$ everywhere.
There are two contributions to the $z$ derivative: first (I), horizontal flux in the plane can wrap
around the field line $L_i$. Secondly (II), $L_i$ can move horizontally about other field lines. 
Now,  $\hat {\theta}_{ij} = {\zhat} \cross \hat { r}_{ij}$ is the angular direction at $\xtwo_j$  about $\xtwo_i$, with $\vec r_{ij} = \xtwo_j - \xtwo_i$. Also, a field line travelling from $(r,\theta,z)$ to $(r+\rmd r,\theta + \rmd \theta,z + \rmd z)$ satisfies the equation
\begin{equation}
 \frac{\rmd  r}{B_r} = \frac{r\rmd \theta}{B_\theta} = \frac{\rmd z}{B_z}.
 \end{equation} 
Thus for a field line at $\xtwo_j = (x, y, z)$, (writing $\vec B_j = B(\xtwo_j)$, etc.) 
\begin{equation}
\left(\frac{\rmd \theta_{ij}}{\rmd z}\right)_I = \frac {B_{\theta j}}{r_{ij} B_{zj}} = \frac{\vec B_j\cdot \zhat \cross \vec r_{ij}}{r^2_{ij} B_{zj}}= \zhat \cdot \frac{\vec r_{ij}\cross\vec B_j }{r^2_{ij} B_{zj}}.
\end{equation}
Next, the field line $L_i$ can move horizontally around other field lines. In this case, 
\begin{equation}
\left(\frac{\rmd \theta_{ij}}{\rmd z}\right)_{II}  = -\zhat \cdot \frac{\vec r_{ij}\cross\vec B_i }{r^2_{ij} B_{zi}} = \frac {\vec B_i }{ B_{zi} }\cdot \frac{\vec r_{ij}\cross  B_{zj} \zhat }{r^2_{ij} B_{zj}}.
\end{equation}
Summing the last two equations, the total change in winding between lines $L_i$ and $L_j$ is
\begin{equation}
\frac{\rmd \theta_{ij}}{\rmd z} = \frac {\vec B_i }{ B_{zi} }\cdot \frac{\vec r_{ij}\cross \vec B_j  }{r^2_{ij} B_{zj}}.
\end{equation}

Let $S_z$ be the horizontal plane at height $z$.
Suppose we add up contributions from all over the plane: we multiply the above expression by$~ B_{zj} \rmd^2 \xtwo_j$ and integrate. Dividing by $2\pi$ to make complete turns into integers,
\begin{equation}
{\deriv{\flh(L_i)}{z}} = \oerpi \frac {\vec B_i }{ B_{zi} }\cdot  \int_{S_z} \frac{\vec r_{ij}\cross \vec B_j  }{r^2_{ij}} ~\rmd^2 \xtwo_j.
\end{equation}

This expression for the field line helicity can be expressed very simply using a special vector potential - the \emph{winding gauge} \citep{prioryeates14}: for a point on field line $L_i$ at
$(x,y, z)$,  
\begin{equation}
\avec^W(x,y, z) = \oerpi   \int_{S_z} \frac{\vec r_{ij}\cross \vec B_j  }{r^2_{ij}} ~\rmd^2 \xtwo_j.
\end{equation}
We have
\begin{equation}
\flh(L_i)= \int \avec^W \cdot \dline = \int \avec^W \cdot \deriv {\mathbf l}{z} ~\rmd z,
\end{equation}
so
\begin{equation}
{\deriv{\flh(L_i)}{z}}= \avec^W \cdot \deriv {\mathbf l}{z}=  \avec^W\cdot \frac {\vec B_i (x,y, z)}{ B_{zi} } .
\end{equation}

\subsubsection{Volume above a plane}
In the previous section we assumed that $B_z > 0$ everywhere. Thus the field lines were braided about each other, and simple winding numbers sufficed to measure pairwise entanglements. If we remove the restriction on $B_z$, then field lines can go up and down.  We will remove the restriction of an upper plane as well in the following discussion, although that is not strictly necessary. Such a domain is illustrated in Figure \ref{fig:winding}b.

If field lines can form loops which go up and down, we can measure topological structure in a few ways. One method is to cut the field lines at any local minima or maxima where $B_z = 0$, then sum winding numbers as before \citep{BergerPrior2006}. The winding gauge will still work in this situation, as the formulae for calculating $\vec A^W$ do not require $B_z>0$.

 A second method, when considering the winding between two loops, consists of examining the angles of the quadrilateral formed by the four foot points of the loops \citep{Berger86,dpb06}. 
Consider the
upper half space $\{z>0\}$, shown in Figure \ref{fig:winding}b. The foot points of 
loops 1 and 2 are labelled
$1^+$, $1^-$, $2^+$, and $2^-$, where $B_z > 0$ at $1^+$ and $2^+$. If the loops cross as seen from above, we assume that loop 1 is the
upper loop.
Consider the quadrilateral $1_+2_+1_-2_-$. Let $\alpha$
and $\beta$ be the angles   at vertices $1_+$ and $1_-$
respectively. Then (from considering the net change of helicity found when bringing the two loops in from infinity) \cite{Berger1984} showed that
 \begin{equation}
 w_{12}=\frac{1}
{2\pi} (\alpha+\beta).\label{h12}
 \end{equation}
 
\subsubsection{Poloidal-toroidal representation} \label{sec:pt}

\newcommand\bt{ {\vec B_{T}} }
\newcommand\bp{ {\vec B_{P}} }
\newcommand\bpn{ \bp\cdot \nhat }
\newcommand\btn{ \bt\cdot \nhat }
\newcommand\at{\vec A_{T}}
\renewcommand\ap{\vec A_{P}}
\newcommand\lapsurf{\Delta_{\parallel}}

The winding gauge employed in the previous sections is equivalent to the vector potential found when employing a poloidal-toroidal representation for the field \citep{Moffatt78, BergerHornig2018}. 

Because the magnetic field $\vec B$ is a three-vector subject to one condition
($\div\vec B =0$) we can often express it in terms of two scalar functions. Here we let $P$ and $T$ be the \emph{poloidal} and \emph{toroidal} functions. Let the horizontal surface Laplacian be $\lapsurf =
\partial^2/\partial_x^2 + \partial^2/\partial_y^2$.
In any plane $z = \textrm{const.}$, $P$ is determined by the vertical flux,
\begin{equation}
\lapsurf P = -B_z
\end{equation}
while $T$ is determined by the vertical current,\begin{equation}
\lapsurf T = -J_z =-(\curl {\vec B})_z.
\end{equation}
We assume $T$, $P$, $B_z$ and $J_z$ all vanish at infinity. Then the surface Laplacians have unique solutions,
\begin{eqnarray}
P(x,y) &=& -\frac{1}{2\pi}\int_{S_z}  B_z(\vec x') \ln |\vec x - \vec x'| \da ';\label{green1}\\
T(x,y) &=& -\frac{1}{2\pi}\int_{S_z}  J_z(\vec x') \ln |\vec x - \vec x' |\da '.\label{green2}
\end{eqnarray}
We can now create a vector potential
\begin{equation}
\vec A = \curl  P\zhat + T \hat z.
\end{equation}
Note that the horizontal components of $\vec A$ only involve the poloidal function $P$. Furthermore, the horizontal divergence of $\vec A$ vanishes, $\grad_{\parallel}\cdot \vec A_{\parallel} = 0$.
One can then employ (\ref{green1}) and (\ref{green2}) to show that this vector potential is identical to the winding gauge, $\vec A^W = \curl  P\zhat + T \hat z$,
provided that the volume considered is infinite in the $x$ and $y$ directions.
We can again employ this vector potential to calculate field line helicities. The total helicity can also be regarded as the net linking of toroidal and poloidal fields \citep{BergerHornig2018},
\begin{equation}
H = 2 \int_\vol \curl(T \hat z) \cdot \curl(P\hat z)\,\rmd ^3 x.
\end{equation}

\subsubsection{Volumes bounded by a sphere} \label{sec:sph}

Spherical boundaries -- such as in Figure \ref{fig:winding}c -- can, for the most part, be treated in the same way as planar boundaries,
with the radial unit vector $\rhat$ replacing the vertical vector $\hat z$. In essence, winding angles become
azimuthal angles: suppose we consider three points $A$, $B$, and $C$ on a sphere. Rotate the spherical coordinates so that $B$ is at the North pole and $A$ is at azimuthal coordinate $\phi = 0$. Then the angle $\angle ABC$ equals the azimuthal coordinate of $C$.

The poloidal and toroidal flux functions are  \citep{Kimura1987}
\begin{eqnarray}
P(\theta, \phi) &=& \frac{1}{4\pi}\int_S  B_r(\vec x') \ln(1-\cos \xi) \da ';\label{green3}\\
T(\theta, \phi) &=& \frac{1}{4\pi}\int_S  J_r(\vec x') \ln (1-\cos \xi)\da '\label{green4},
\end{eqnarray}
where
\begin{equation}
 \qquad \cos \xi = \cos\theta\cos \theta ' + \sin\theta\sin\theta '\cos(\phi-\phi')
\end{equation}
is the spherical distance between $(\theta, \phi)$ and  $(\theta', \phi')$. With these functions,
\begin{equation}
\vec A = \grad  P\cross \rhat + T \hat r;\qquad 
\vec B = \curl(\grad  P\cross \rhat) + \grad  T\cross \rhat.
\end{equation}

Some care must be taken in spherical geometries. One must make sure that no magnetic monopoles are hiding inside the sphere -- i.e. the net flux through the sphere must vanish (similarly for the net current). 

Also, in the special case of a spherical shell geometry where flux enters at the inner sphere and exits at the outer sphere (as in Figure \ref{fig:winding}c) the separation of helicity into \emph{self helicity} and \emph{mutual helicity} can be ambiguous \citep{campbell2014}. Self helicity measures the twist and writhe of individual tubes, while mutual helicity measures linking  and intertwining between tubes. This will not affect the field line helicity, however.

\newcommand{\ncurl}{\mathcal{D} }
\newcommand{\ncurlinv}{\mathcal{D}^{-1}}
\newcommand{\asurf}{{\vec A}_{\parallel}}

\subsubsection{More general volumes} \label{sec:gen}

For volumes bounded by planes or spheres there are natural definitions of angle and winding angle depending only on the Euclidean metric. The winding gauge, or equivalently the poloidal-toroidal vector potential, captures this natural definition. We could also consider a volume with boundaries consisting of a bottom plane and side boundaries, where flux only crosses the bottom plane. Such a volume could represent a closed active region. We could choose winding angles without regard to the side boundaries, using the same winding gauge. An alternative will be described below, when using relative measures of helicity.

Next consider a volume bounded by one or two simply connected surfaces, but which lack the symmetry of planes or spheres. An example is shown in Figure \ref{fig:winding}d. Here definitions of angle are less obvious. However, we can still employ a generalization of the poloidal-toroidal representation \citep{BergerHornig2018} to define the winding gauge.

Let our volume $\vol$ be sliced into nested surfaces. Employ coordinates $(u,v,w)$ where $w = \textrm{const.}$ labels one of the nested surfaces. The toroidal field (which is responsible for the current perpendicular to surfaces of constant $w$) can be written in terms of a toroidal function $T$ as before:
\begin{equation}
\vec B_T = \curl \vec A_T; \qquad \vec A_T = T\grad w.
\end{equation}

However, the poloidal field is more complicated. Let the normal component of the curl be given by the operator $\ncurl$,
\begin{equation}
\ncurl \vec V = \nhat\cdot \curl \vec V.
\end{equation}
This operator has an inverse $\ncurlinv$. To make the inverse unique, we require that the inverse normal curl gives a divergence free vector field parallel to the surface. Thus
\begin{equation}
\ncurlinv f \cdot \nhat = 0; \qquad \div \ncurlinv f = 0.
\end{equation}

Let $B_n$ be the normal magnetic field. Ordinarily, we would write the poloidal field as
\begin{equation}
\tilde {\vec A}_P = \ncurlinv B_n; \qquad \vec B_P = \curl \tilde {\vec A}_P.
\end{equation}
However, without spherical or planar symmetry this magnetic field may have nonzero normal current $J_n$ -- but the toroidal field has already taken care of $J_n$. Thus \citep{BergerHornig2018} we must add a \emph{shape} term
\begin{equation}
\vec A_S = T_S\grad w
\end{equation}
where
\begin{equation}
\curl\curl A_s \cdot \nhat = - \curl\curl \tilde {\vec A}_P \cdot \nhat.
\end{equation}
Putting it all together, the generalized winding gauge becomes
\begin{equation}
\vec A^W = \ncurlinv B_n + (T+T_S)\grad w.
\end{equation}
 
\subsection{Relative field line helicity} \label{sec:rflh}

In the previous sections, helicity and field line helicity have been defined in terms of geometrical quantities such as winding and linking. The helicity of open fields in a volume $\vol$ can also be defined in terms of how much the magnetic structure within $\vol$ contributes to the helicity of all space \citep{BergerField}. Let the field inside $\vol$ be called $\vec B_{\vol}$. Also consider a potential (zero current) field $\vec P_{\vol}$ inside $\vol$ with the same boundary conditions $\vec B_{\vol}\dotn = \vec P_{\vol}\dotn$. A potential field minimizes the magnetic energy in the volume subject to the constraint of the given boundary flux distribution. Hence potential fields can be said to have the minimum structure (more accurately, their structure only depends on the boundary conditions, so they add zero \emph{additional} structure to the field). 

Here we compare the total helicity of all space with the helicity that would be obtained if, inside $\vol$, $\vec B_{\vol}$ were replaced by $\vec P_{\vol}$. All integrals are gauge invariant, so the difference between the two helicities will also be gauge invariant. Let space external to $\vol$ be $\vol'$. Then the relative helicity $H_R(\vol)$ is
\begin{equation}
H_R(\vol)= H(B_\vol, B_{\vol'}) - H(P_\vol, B_{\vol'}) . 
\end{equation}
It is important to note that the calculation of $H_R(\vol)$ does not require knowledge of
the external field $B_{\vol'}$ \citep{BergerField,FinnAntonsen1985}.

For symmetric volumes (i.e. boundaries consisting of a plane or parallel planes or one or two concentric spheres), the topological and relative definitions give the same answers \citep{BergerField,BergerHornig2018}. For more complicated geometries they may differ. For example, suppose the volume $\vol$ consists of a helical tube rising between parallel planes. Suppose we place a potential field both inside and outside of $\vol$. Then the helicity of all space will be non-zero because of the helical structure. Now add some axial current inside $\vol$ so that the field has extra twist. Then the relative helicity inside $\vol$ will only measure this extra twist, because the helicity due to the axis writhe is subtracted off with the potential term.

The relative helicity of a potential field is, by definition, zero. Suppose again that the boundary is planar or spherical. Then the topological helicity (summing winding numbers as in previous sections) will be zero as well: the potential field is purely poloidal so there is no linkage between toroidal and poloidal components. However, for a potential field, the  field line helicities of individual field lines may not be zero, even if they sum to zero \citep{Yeates2020}. These potential field line helicities reflect irregularities and lack of mirror symmetry in the distribution of boundary flux \citep{Bourdin2018}. Thus field line helicity of the potential field can help to characterize magnetic distributions, for example in the solar photosphere. 

This raises a question, however: if there are currents inside a volume, how do they change the field line helicities from their minimal (potential field) values? We need to subtract off the potential field line helicities to remove the contributions from the boundary flux distribution.
Thus, we can adopt the relative helicity viewpoint for field line helicity as well \citep{YeatesPage2018,Moraitis2019,Moraitis2021}. 
First, to avoid gauge ambiguity, we employ the same gauge conditions as apply in the poloidal-toroidal formulation. Let the potential field be $\vec P$ with vector potential $\vec A_P$. In particular, for the components of the vector potentials parallel to the boundary, we have 
\begin{eqnarray}
\grad \cdot\vec A_{\parallel} &=& \grad \cdot \vec A_{P \parallel} = 0; \label{eqn:dapar}\\
\vec A_{\parallel} &=&\vec A_{P\parallel} . \label{eqn:apar}
\end{eqnarray}
It can be shown \citep{YeatesPage2018} that this choice of gauge for $\vec A_P$ minimises the boundary integral $\oint_{S}|\vec A_P\cross \nhat|^2 ~ \rmd ^2 x$. 
Next let $\alpha_+$ be the positive endpoint of a field line. The field lines through $\alpha_+$ for $\vec B$ and $\vec P$ will in general have different trajectories, and will connect to different exit points $\alpha_-$ and $\alpha_{P-}$. We can now take the difference in field line helicities
\begin{equation}
\flh^+(\alpha_+) = \int_{\alpha_+}^{\alpha_-} \vec A \cdot \dl - \int_{\alpha_+}^{\alpha_{P-}} \vec A \cdot \dl.
\end{equation}
We could also do this for the negative endpoints. Starting at the same negative endpoint $\alpha_-$, we could track the field lines backwards to either $\alpha_+$ or $\alpha_{P+}$. In this case, we can write
\begin{equation}
\flh^-(\alpha_-) = \int_{\alpha_+}^{\alpha_-} \vec A \cdot \dl - \int_{\alpha_{P+}}^{\alpha_{P-}} \vec A \cdot \dl.
\end{equation}
Finally, we can take the average of the two (for the endpoints $\alpha_+$ and $\alpha_-$ of the true field line $\vec B$ to find 
\begin{equation}
\flh^0 = \half (\flh^+(\alpha_+)+\flh^-(\alpha_-)).
\end{equation}
This is the measure of \emph{relative field line helicity} adopted by \citet{Moraitis2019}, although those authors use a slightly different gauge in their computations.

\section{Ideal evolution} \label{sec:ideal}

Just like the total magnetic helicity, the field line helicities within a given domain $\vol$ may change in two ways: by (i) boundary motions or (ii) non-ideal processes in the interior. In this section we show how to derive an evolution equation for case (i), while case (ii) is deferred to Section \ref{sec:nonideal}. We will give some examples where these have been applied, though much remains to be explored.

If $\vec B$ evolves ideally in some volume $\vol$, so that
\begin{equation}
\pder{\vec B}{t} = \nabla\times\big(\vec u\times\vec B\big),
\label{eqn:indB}
\end{equation}
then it is well known that magnetic field lines retain their identity: if two fluid elements lie on the same magnetic field line at some initial time, they will continue to do so at all later times. Moreover, by Alfv\'en's Theorem, field lines cannot change their mutual linkage. It follows immediately that $\flh(L)$ is preserved for every \emph{closed} field line $L$ in $\vol$. Moreover, any functional of $\flh$ will also be preserved, including not only the total helicity
\begin{equation}
H = \int_\vol\avec\cdot\vec B\dv = \int_L\flh(L)\mathrm{d}\Phi
\end{equation}
but also the helicity of any subregion composed of closed field lines, as well as the unsigned helicity defined by
\begin{equation}
\overline{H} = \int_L|\flh(L)|\mathrm{d}\Phi.
\end{equation}

\begin{figure}
\centering
\includegraphics[width=0.85\textwidth]{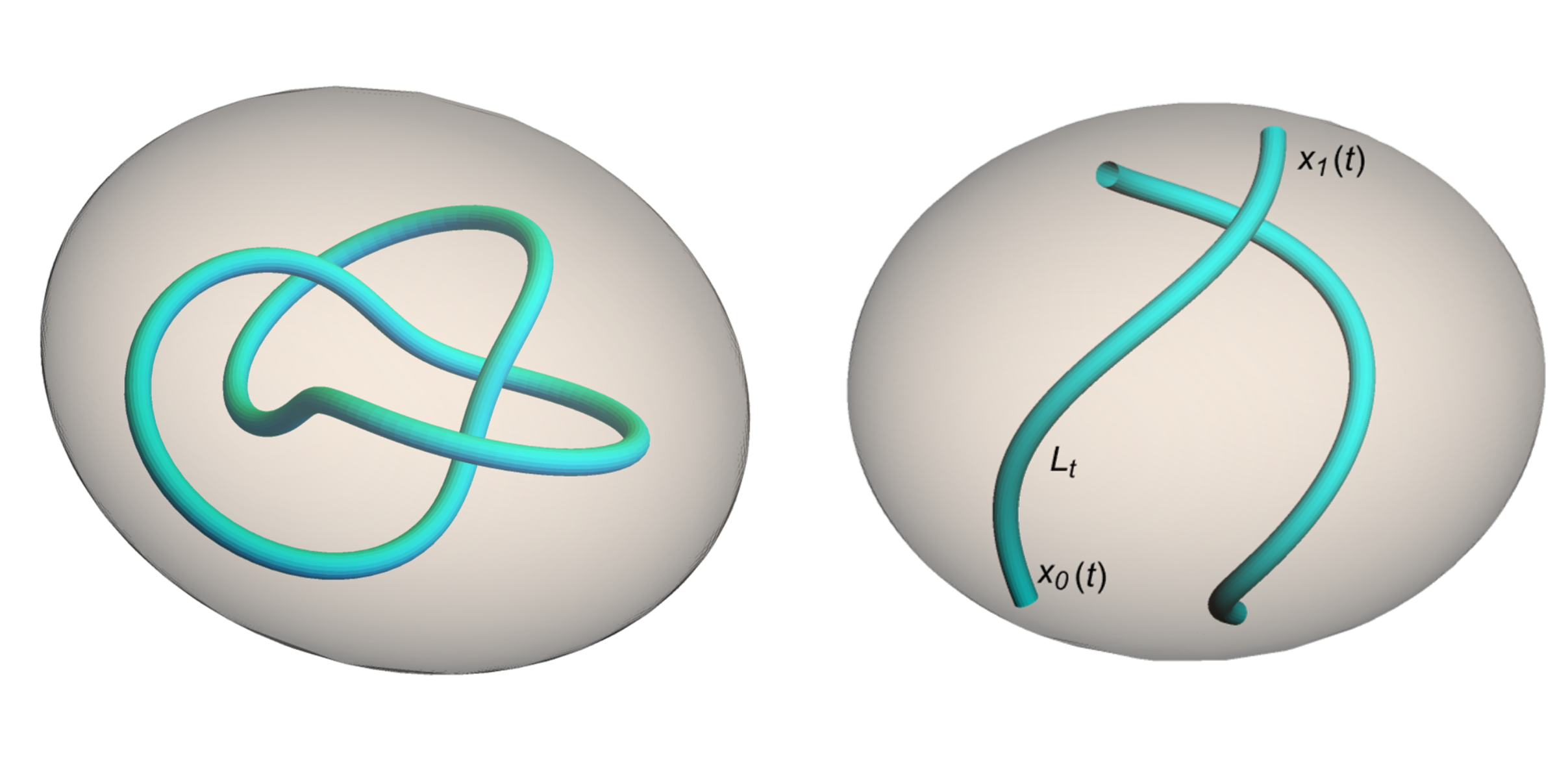}
\caption{Examples of closed (left) and open (right) field lines in a domain $V$ with connected boundary $\partial \vol$. Footpoints of the open field line $L_t$ are denoted $x_0(t)$ where $\vec B\dotn < 0$ and $x_1(t)$ where $\vec B\dotn > 0$. }
\label{fig:openclosed}
\end{figure}

To inject or remove field line helicity by a continuous ideal evolution therefore requires \emph{open} field lines where $B_n\equiv\vec B\dotn\neq 0$ along with $\vec u\neq\vec 0$ on the boundary $S$, as illustrated in Figure \ref{fig:openclosed}(b). In this case, the fundamental issue of gauge choice is important.
A physically meaningful approach is to fix something like the winding or poloidal-toroidal gauge for ${\vec A}$ at all times. Consider an open field line $L_t$, where the subscript $t$ denotes that this line moves with the fluid. To derive the evolution equation for $\flh(L_t)$, note first that uncurling \eqref{eqn:indB} gives
\begin{equation}
\pder{\avec}{t} = \vec u\times\vec B + \nabla\phi
\label{eqn:indA}
\end{equation}
for some scalar function $\phi(\vec x,t)$ that depends on the chosen gauge of $\avec$ and will generally be non-zero for the poloidal-toroidal gauge.

Now let $\vol_{t,\epsilon}$ be an infinitesimal flux tube surrounding $L_t$ and moving with the fluid. Let $\Phi(\vol_{t,\epsilon})$ and $h(\vol_{t,\epsilon})=\int_{\vol_{t,\epsilon}}\avec\cdot\vec B\dv$ be the magnetic flux and helicity of this flux tube. Then the rate of change of field line helicity is 
\begin{align}
 \deriv{}{t}{\flh(L_t)} &= \pder{}{t}\lim_{\epsilon\to 0}\frac{h(\vol_{t,\epsilon})}{\Phi(\vol_{t,\epsilon})}\\
&= \lim_{\epsilon\to 0}\frac{1}{\Phi(\vol_{t,\epsilon})}\deriv{h(\vol_{t,\epsilon})}{t}. \label{eqn:dalim}
 \end{align}
By the Reynolds' Transport Theorem, since $\vol_{t,\epsilon}$ is a material volume, we have
\begin{align}
\deriv{}{t}{h(\vol_{t,\epsilon})} &= \int_{\vol_{t,\epsilon}}\pder{}{t}(\avec\cdot\vec B)\rmd^3x + \oint_{S_{t,\epsilon}}(\avec\cdot\vec B)\vec u\cdot\nhat\da,
\end{align}
which using \eqref{eqn:indA} reduces to
\begin{align}
\deriv{}{t}{h(\vol_{t,\epsilon})} &= \oint_{S_{t,\epsilon}}\big(\phi + \avec\cdot\vec u\big)\vec B\cdot\nhat\da.
\end{align}
This is just the standard expression for evolution of magnetic helicity in a material volume (e.g. \cite{moffatt2019self} p.61). Since $\vol_{t,\epsilon}$ is a magnetic flux tube, $\vec B\cdot\nhat$ is non-zero only on the two portions of the tube boundary $S_{t,\epsilon}$ that coincide with the boundary of $\vol$ -- in other words, the ends of the tube. So substituting into \eqref{eqn:dalim} and taking the limit, noting that $\vec B\cdot\nhat \to \Phi(\vol_{t,\epsilon})$ at both ends of the tube, gives
\begin{align}
\deriv{}{t}{\flh(L_t)} &= \Big[\phi + \avec\cdot\vec u\Big]_{x_0(t)}^{x_1(t)}, \label{eqn:dadt}
\end{align}
where $x_0(t)$ and $x_1(t)$ denote the end-points of $L_t$ on the boundary of $\vol$.

The expression \eqref{eqn:dadt} shows that, even though field lines $L_t$ can be uniquely tracked in time during an ideal evolution, their field line helicities $\flh(L_t)$ depend on the chosen gauge of $\avec$ on the boundary over time, since this changes both $\avec\cdot\vec u$ and also $\phi$, the latter being defined for a given choice of $\avec$ through equation \eqref{eqn:indA}. For meaningful results this gauge should be chosen objectively -- such as the poloidal-toroidal gauge -- rather than for computational convenience. Indeed, it is always possible to ``cancel'' the effect of an ideal evolution by choosing $\phi = -\avec\cdot\vec u$ in equation \eqref{eqn:indA}, so that $\flh$ remains invariant for all field lines,  but this is a highly artificial choice; in effect, it corresponds to measuring winding with respect to a frame moving with the boundary motions. Of course, if there are no boundary motions, then $\flh$ will be invariant in time for all field lines provided that $\phi$ on the boundary is not varied. In particular, if the boundary is a connected surface then it is sufficient to fix ${\vec A}\times\nhat$. In cases such as the volume between two spheres, one must also fix the integral of ${\vec A}$ along a line between the two boundaries.

When there are boundary motions, the winding gauge makes it possible to express the evolution of $\flh$ in terms of the mean angular motions of field line endpoints around one another \citep{1988A&A...201..355B, 2021GApFD.115...85M}. In the restricted case where there are boundary motions but they preserve the boundary distribution $B\dotn$, and in addition the magnetic field has a ``simple topology'' (meaning that the mapping from positive to negative endpoints is continuous), \citet{2018FlDyR..50a1408A} derived the alternative formula
\begin{equation}
\deriv{}{t}{\flh(L_t)} = \left[ U\pder{\zeta}{U} - \zeta\right]_{x_0(t)}^{x_1(t)}.
\label{eqn:aly}
\end{equation}
Here the boundary motions have been written as $\vec u = (\nhat\times\nabla\zeta)/{\vec B}\cdot\nhat$, and the gauge of ${\vec A}$ has been chosen in a particular way. Specifically, the parallel components of $\avec$ on the boundary are set by equation \eqref{eqn:apar}, but with $\avec_P=U\nabla V$ where $U$ and $V$ are Euler potentials for the reference field $\vec P=\nabla U\times\nabla V$. Formula \eqref{eqn:aly} may be shown to be equivalent to \eqref{eqn:dadt} for this gauge choice.

\subsection{Simple examples}

The simplest examples are obtained by considering a magnetic field between two planes $z=0$ and $z=1$, on which ${\vec B}\dotn=B_z=1$, and applying axisymmetric boundary motions on the upper plane. In this situation, the potential reference field is $\vec P=\zhat$ at all times, with poloidal-toroidal vector potential $\avec_P = (r/2)\thetahat$. We fix the gauge of $\flh$ by setting $\avec\times\zhat=\avec_P\times\zhat$ on $z=0$ and $z=1$, and by the additional condition (because the boundary has two disconnected components $z=0$ and $z=1$) that $\int_L\avec\cdot\dl=\int_L\avec_P\cdot\dl=0$ for the vertical line $L$ at $r=0$. This gives us $\flh$ in the winding, or equivalently poloidal-toroidal gauge (the two are equivalent in this volume).

\begin{figure}
\centering
\includegraphics[height=0.35\textwidth]{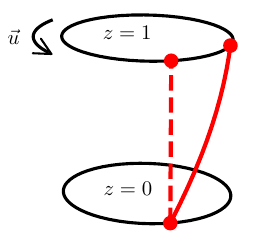}
\includegraphics[height=0.35\textwidth]{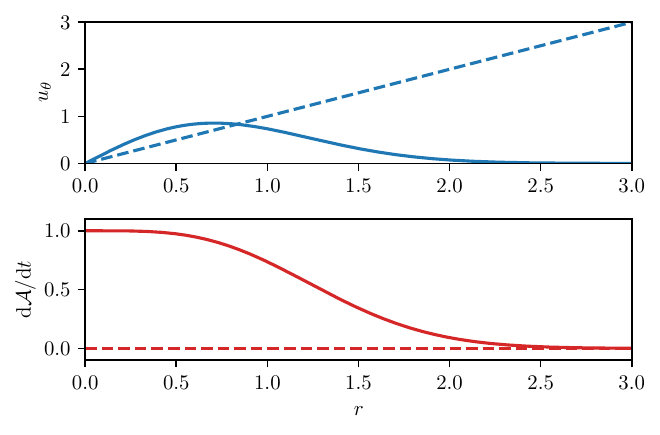}
\caption{Simple example where the boundary has two parallel components, one of which is rotated with respect to the other. The graphs show the azimuthal velocity (top) and rate of generation of field line helicity (bottom), for the two examples in the text (dashed and solid lines).}
\label{fig:twisting}
\end{figure}

Suppose firstly that we apply a rigid rotation of the upper boundary with respect to the lower one, given by $\vec u = rz\thetahat$. In this case $\vec u\times\vec B = u_\theta B_z\rhat = rz\rhat$, so maintaining the gauge condition $\avec\times\zhat=\avec_P\times\zhat$ requires, according to \eqref{eqn:indA}, that
\begin{equation}
\pder{\phi}{r} = -rz \quad \implies \quad \phi = -\frac{r^2z}{2}.
\end{equation}
Thus  for a field line at radius $r$, equation \eqref{eqn:dadt} gives
\begin{equation}
\deriv{}{t}{\flh(r)} = \left[ -\frac{r^2z}{2} + \left(\frac{r}{2}\right)rz\right]_{z=0}^{z=1} = 0.
\label{eqn:twist}
\end{equation}
This shows that a rigid rotation generates no field line helicity (as noted by \citet{2021arXiv210801346Y}). In effect, all of the field lines rotate together so that none of them acquires any twist with respect to the others. This example is shown by the dashed lines in the right-hand panels of Fig. \ref{fig:twisting}. 

For an example with non-trivial injection of field line helicity, we can apply a localized rotation  $\vec u = 2r\mathrm{e}^{-r^2}\phihat$,  for which a similar computation leads to a localized patch of field line helicity,
\begin{equation}
\deriv{}{t}{\flh(r)} = (r^2 + 1)\mathrm{e}^{-r^2}.
\label{eqn:gausstwist}
\end{equation}
This is shown by the solid lines in the right-hand panels of Fig. \ref{fig:twisting}. One obtains the same $\flh$ pattern in the topologically equivalent situation of a toroidal flux ring that is localized in $z$, as shown by \citet{2014JPhCS.544a2002Y}. Indeed, the initial condition in Fig. \ref{fig:braidsims} (later) is produced by the superposition of six such flux rings, offset so that the field lines follow a tangled pattern.

\begin{figure}
\centering
\includegraphics[width=0.6\textwidth]{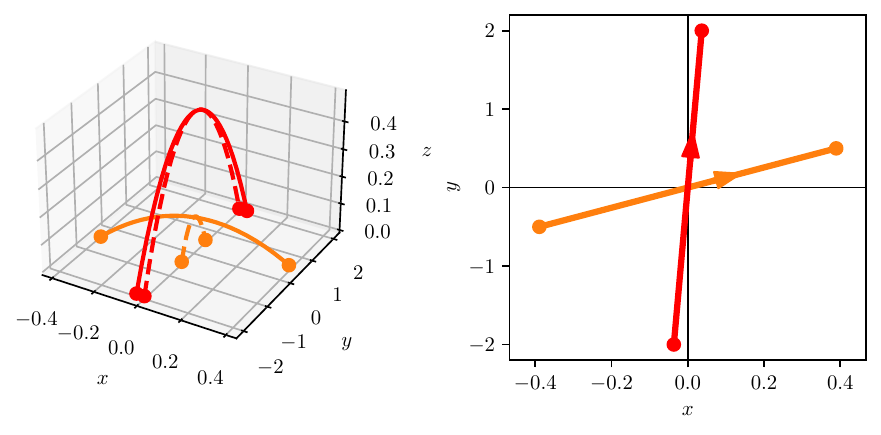}
\includegraphics[width=0.38\textwidth]{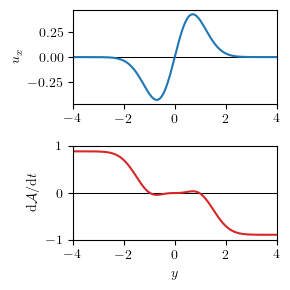}
\caption{Simple example of an initial potential ``arcade'' being sheared by a flow on the lower boundary. Although they are largely unsheared, the overlying field lines acquire negative field line helicity, as may be seen from their negative crossings with the sheared field lines beneath.}
\label{fig:shearing}
\end{figure}

A final analytical example demonstrates how field line helicity may be injected into a magnetic field defined on the half-space $z>0$ through shearing motions on the lower boundary $z=0$, as opposed to twisting. This models the shearing of an arcade of coronal magnetic loops in the Sun's atmosphere. We suppose that the arcade is initially current-free and (for simplicity) invariant in $x$, with $\vec B = \vec P =  z\yhat - y\zhat$, the solar surface being the plane $z=0$ and the volume $V$ with $z>0$ representing the corona. A suitable poloidal-toroidal reference vector potential is $\avec_P = (y^2/2 + z^2/2)\xhat$. Suppose we apply the constant shearing velocity $\vec u=y\mathrm{e}^{-y^2}\xhat$, localised near the polarity inversion line $y=0$.  Here  $\vec u\times\vec B  = -u_xB_z\yhat + u_x B_y\zhat$, so requiring that $\nhat\times\avec = \nhat\times\avec_P$ on the boundary $z=0$ implies that
\begin{equation}
\pder{\phi}{y} = u_xB_z = -y^2\mathrm{e}^{-y^2} \quad \implies \quad \phi = -\frac14\Big(\sqrt{\pi}\,\mathrm{erf}(y) - 2y\mathrm{e}^{-y^2}\Big).
\end{equation}
Substituting into \eqref{eqn:dadt} and evaluating on the field line rooted at $y>0$ then gives
\begin{equation}
\deriv{}{t}{\flh(y)} =  (y+y^3)\mathrm{e}^{-y^2}-\frac{\sqrt{\pi}}{2}\,\mathrm{erf}(y).
\label{eqn:sheared}
\end{equation}
This profile is shown in Fig. \ref{fig:shearing}. The first term vanishes on the unsheared field lines at large $y$. But the second term does not. This demonstrates the non-local nature of helicity: the field lines in the overlying arcade gain a field line helicity because the core of the arcade is sheared. Effectively there is a magnetic flux passing through them. Indeed, this was precisely the way that \citet{Antiochos1987} proposed to define field line helicity, as mentioned earlier. Notice here that the sign of the $\mathrm{erf}(y)$ term is negative, consistent with Stokes Theorem and the direction of the sheared field relative to the orientation of the overlying arcade. One could also infer this sign by looking at the arcade from above and observing a negative crossing (Fig. \ref{fig:shearing}).

\subsection{Application to the global solar corona} \label{sec:global}

To first approximation, the magnetic field in the Sun's atmosphere evolves ideally -- in response to emergence of new magnetic active regions from inside the Sun, decay of these strong magnetic fields due to convective shredding, and transport of the resulting magnetic flux by large-scale motions such as the Sun's differential rotation. During these processes, magnetic helicity builds up in the corona and is ejected into the heliosphere. Field line helicity offers the exciting prospect of a localised measure for studying where this helicity is located within the corona.

\begin{figure}
\centering
\includegraphics[width=0.32\textwidth]{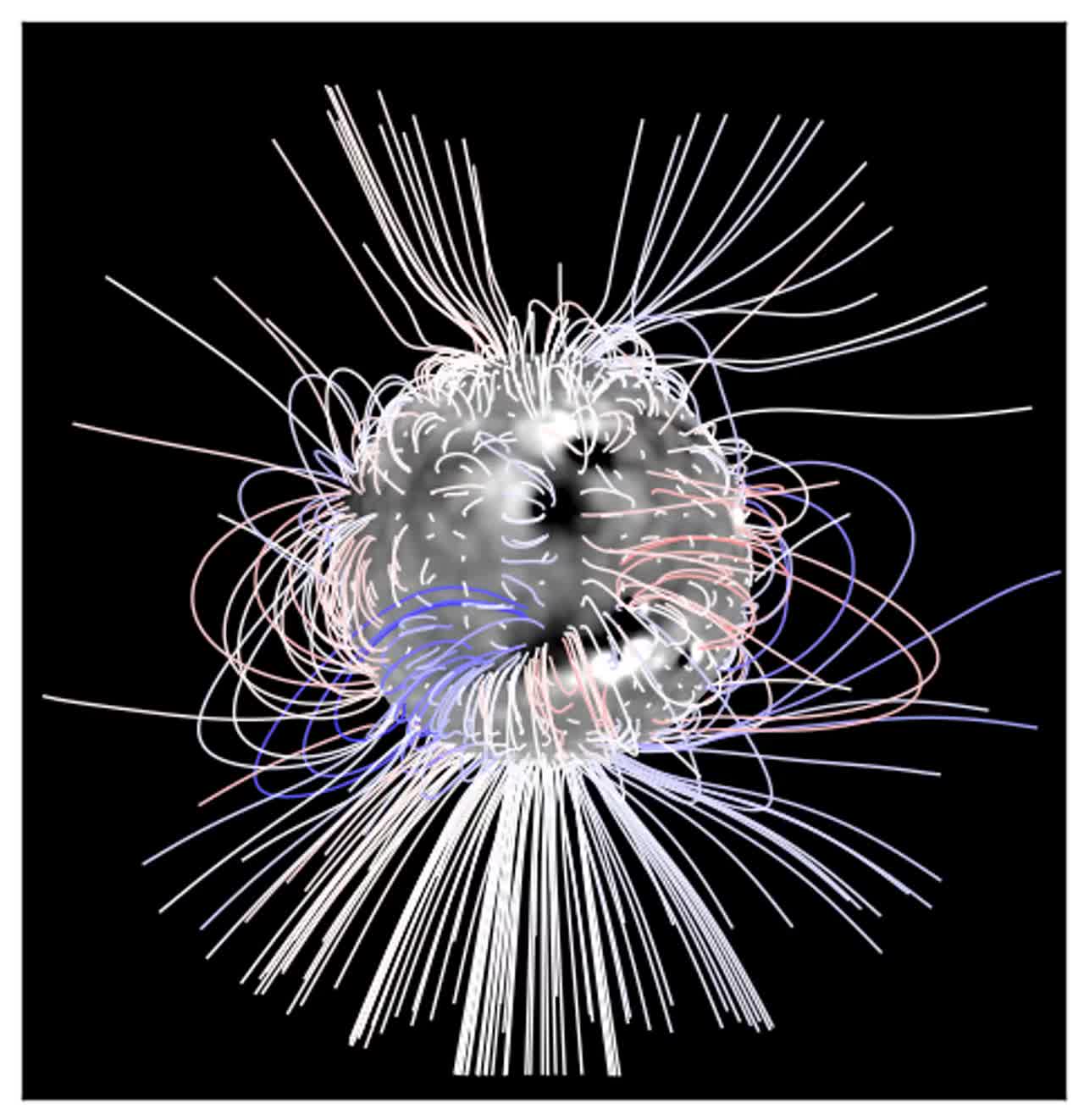}
\includegraphics[width=0.32\textwidth]{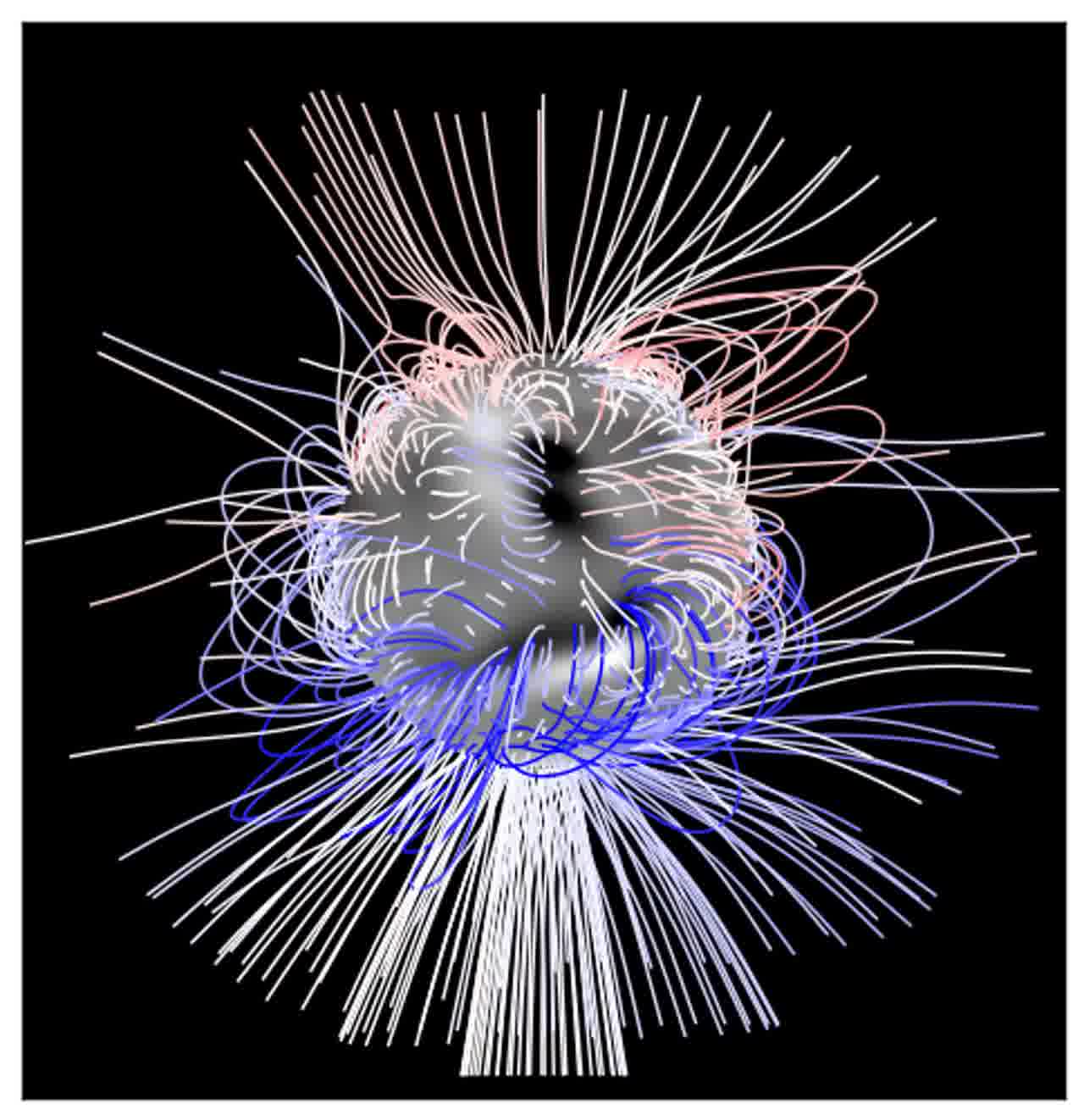}
\includegraphics[width=0.32\textwidth]{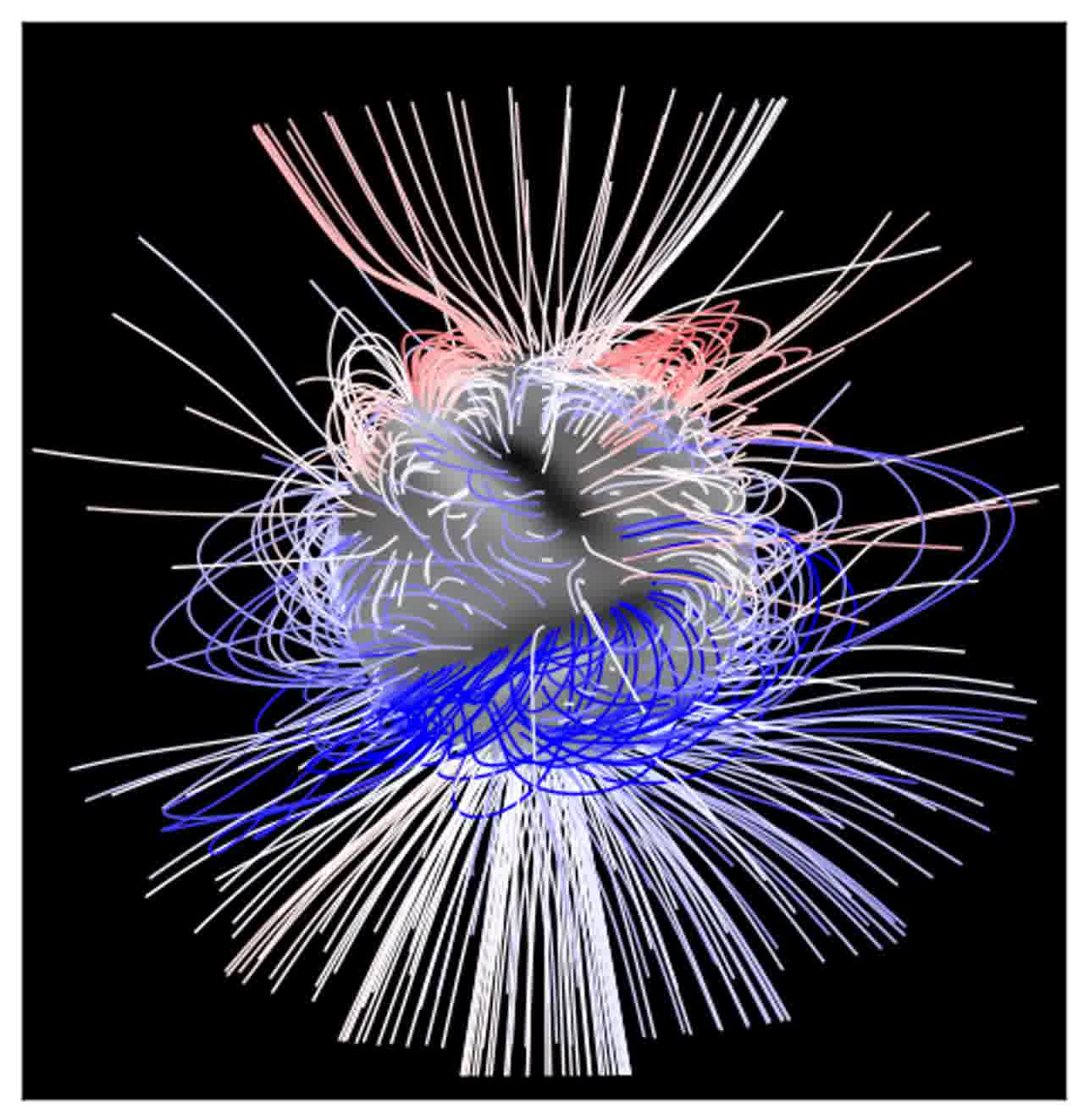}
\caption{Evolution of field line helicity in a magneto-frictional model of the solar corona as an initial potential field (a) is sheared by the Sun's differential rotation for (b) 25 days and (c) 50 days. Field lines are coloured by their field line helicity in poloidal-toroidal gauge, with red for positive and blue for negative (for details of the simulation, see \citep{2016A&A...594A..98Y}).}
\label{fig:global}
\end{figure}

\citet{2016A&A...594A..98Y} used a global magneto-frictional model in a spherical shell to study how field line helicity evolves in the corona, in response to evolution of the solar surface magnetic field (Fig.~\ref{fig:global}). In the simulation shown, active region emergence is neglected, so that the dominant injection of helicity is shearing by differential rotation of field line footpoints on the solar surface. The behaviour of field line helicity is found to be different on open and closed magnetic field lines. On open field lines (meaning in this context that one field line endpoint is on the outer boundary), injected field line helicity is lost at a steady rate through the outer boundary by relaxation of the field line. By contrast, closed field lines, with both endpoints on the solar surface, can store field line helicity -- like in the simple arcade example in Fig.~\ref{fig:shearing}.

This storage of helicity on closed field lines is important, because eventually these form twisted flux ropes that lose equilibrium and erupt, leading to bursts of helicity flux output and believed to explain the origin of coronal mass ejections. \citet{2017ApJ...846..106L} studied these eruptions in the magneto-frictional model, and used field line helicity as a diagnostic tool to define flux ropes in the first place. The overall magnetic flux and helicity content of these structures was found comparable to that estimated in observations of interplanetary magnetic clouds. Recently, \citet{2021SoPh..296..109B} have used field line helicity to show that episodic losses of helicity in the magneto-frictional model come not only from the eruption of flux ropes formed along polarity inversion lines in the low corona, but also from a second type of eruption generated in the overlying streamers. Similar eruptions are known in MHD simulations \citep{1995ApJ...438L..45L} and have been suggested as a possible explanation for so-called stealth CMEs that lack an obvious low-coronal source \citep{2016JGRA..12110677L}. Many unanswered questions remain, not least the role of active regions in the global helicity balance, or the possibility of long-term storage of their helicity in the corona. Field line helicity will greatly facilitate these investigations.

\subsection{Application to solar active regions} \label{sec:ar}

\begin{figure}
\centering
\includegraphics[width=\textwidth]{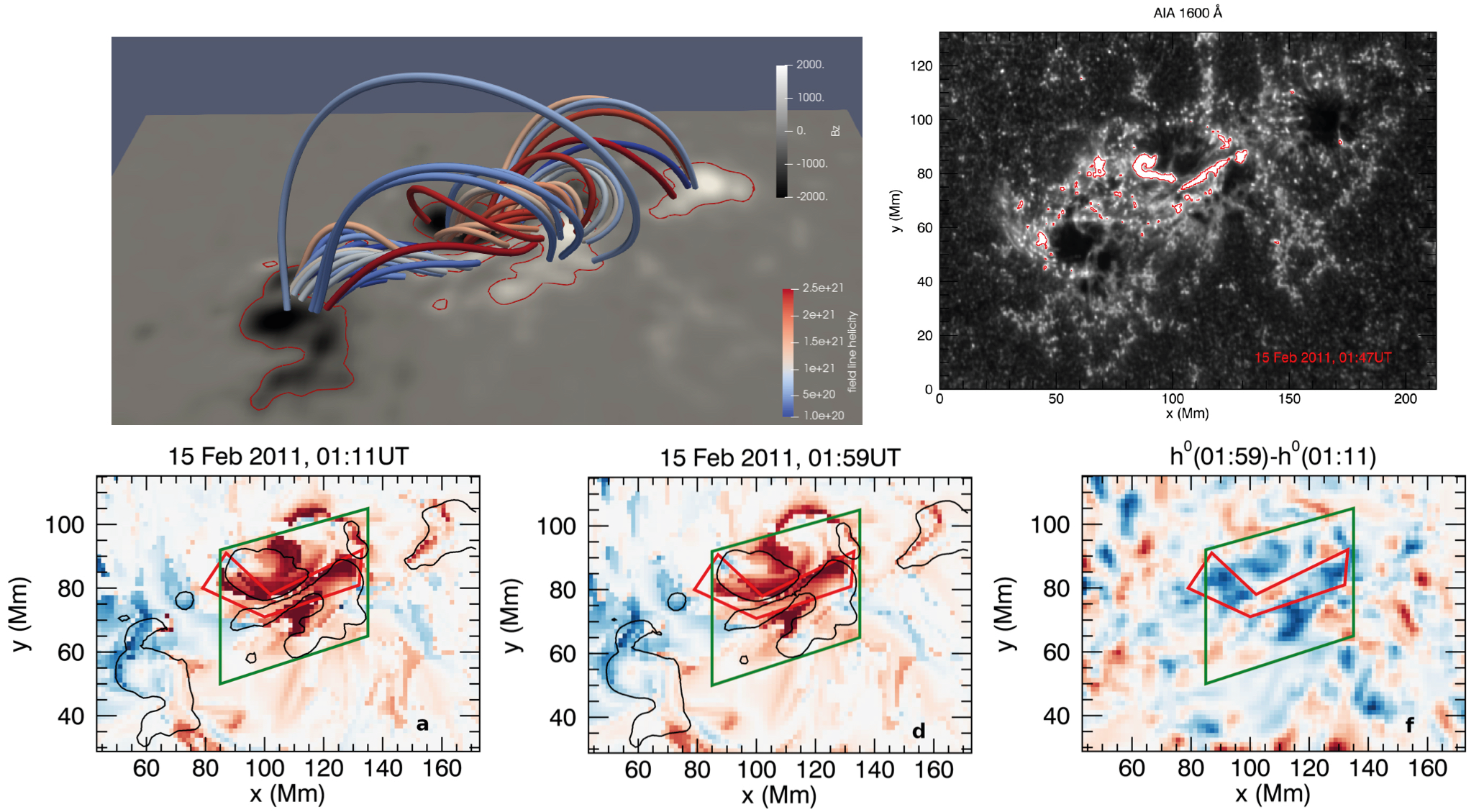}
\caption{Field line helicity in a nonlinear force-free model of active region NOAA 11158, from \citet{Moraitis2021}. Panel (a) shows the field lines coloured by (relative) field line helicity in the extrapolation at 01:11UT, while panel (b) shows an image of EUV emission from SDO/AIA during the X-class flare at 01:47UT. Panels (c) and (d) show (relative) field line helicity before and after the flare (blue/red), while panel (d) shows their difference. Credit: K. Moraitis et al., A\&A, 649, A107, 2021, reproduced with permission \copyright{} ESO.}
\label{fig:moraitis}
\end{figure}

Field line helicity also offers the possibility to identify locations of helicity storage on a smaller scale, within individual active regions, provided three-dimensional magnetic field models are available. Even in a potential field model, a simple bipolar active region can have non-zero field line helicity if its field lines are linked with the overlying background field \citep{Yeates2020}. But much larger values of field line helicity are expected in more realistic current-carrying models of active regions. This has been confirmed by \citet{Moraitis2019} both in idealized MHD models and in extrapolations. For example, \citet{Moraitis2021} have computed field line helicity in nonlinear force-free extrapolations of a real active region NOAA 11158 (Fig.~\ref{fig:moraitis}).

Figure \ref{fig:moraitis} shows a highly sheared magnetic field in the core of the active region, which has positive field line helicity. The authors found that during a flare the region lost 25\% of its relative magnetic helicity. Comparing the field line helicity between the two extrapolations shown at 01:11UT and 01:59UT reveals that the decrease in helicity took place within the same region (the green box) where emission from a large X-class solar flare was observed in EUV. This supports the idea that there was indeed a decrease in helicity associated with the flare. It is worth noting that \citet{Moraitis2021} uses relative field line helicity (Section \ref{sec:rflh}). However, they show that using the ordinary field line helicity (in poloidal-toroidal gauge) leads to the same qualitative conclusions in this example, albeit with lower values. 

\section{Non-ideal evolution} \label{sec:nonideal}

Suppose \eqref{eqn:indB} is generalized to
\begin{equation}
\pder{\vec B}{t} = \nabla\times\big(\vec u\times\vec B\big) - \nabla\times\vec N,
\label{eqn:indBN}
\end{equation}
where $\vec N = \vec E + \vec u\times\vec B$ represents some non-ideal term in Ohm's Law. A common example would be $\vec N=\eta\vec J$ corresponding to resistive MHD. It is well-known that the total helicity is no longer conserved but can be dissipated  within the volume when $\vec N\cdot\vec B\neq 0$. In this section, we show how -- for magnetic fields of simple topology without null points -- equations \eqref{eqn:dadt} and \eqref{eqn:aly} have been generalized to this non-ideal case.  For magnetic fields of more complex topology -- for example the solar coronal examples in Sections \ref{sec:global} and \ref{sec:ar} -- evolution equations have not yet been derived explicitly. Indeed differential equations may not be appropriate since the distribution of $\mathcal{A}$ is discontinuous across magnetic separatrices between different connectivity domains. Reconnection can transport $\mathcal{A}$ across these separatrices, but its evolution in such a situation remains to be studied in detail.

There is an important caveat to what follows: field line helicity can hold physical significance in a non-ideal evolution only if the magnetic field lines themselves  retain sufficient identity over time for their topology to play a physical role. In practice, this means that the magnetic Reynolds number ($\mathrm{Rm}$) must be sufficiently large, or equivalently the (effective) resistivity must be sufficiently small. There is no precise threshold for this, but it is already accessible for the parameters achievable in numerical simulations, as will be illustrated in Section \ref{sec:relax}.

\subsection{Evolution equation for non-null magnetic fields}

The trick for generalizing \eqref{eqn:dadt} or \eqref{eqn:aly} to non-ideal evolution in the non-null case is to decompose $\vec N$ into parallel and perpendicular parts \citep{2011PhPl...18j2118Y}, writing
\begin{equation}
\vec N = -\vec v\times\vec B + \nabla\psi.
\label{eqn:ndecomp}
\end{equation} 
When the magnetic field has a simple topology, this decomposition exists globally with $\vec v$ and $\psi$ continuous throughout $V$, though they are not unique (as we shall discuss below). 
Substituting this decomposition into \eqref{eqn:indBN} shows that, in a non-null magnetic field, we can write
\begin{equation}
\pder{\vec B}{t} = \nabla\times\big(\vec w\times\vec B\big), \quad \textrm{where} \quad \vec w=\vec u+\vec v,
\end{equation}
which shows that the magnetic field is still frozen in, but to the so-called \emph{field line (transport) velocity} $\vec w$ rather than the plasma velocity $\vec u$. Thus in a non-ideal evolution where $\vec v\neq\vec 0$, the field lines slip at some velocity $\vec v$ through the plasma \citep{1958AnPhy...3..347N, 1992JGR....97.1521P, 2006SoPh..238..347A,2010pspa.book.....S}. Uncurling, we have
\begin{equation}
\pder{\avec}{t} = \vec w\times\vec B + \nabla(\phi - \psi).
\end{equation}
Thus, if we identify the same field line $L_t$ at different times by the fact that it is frozen in to the flow of $\vec w$, then a similar argument to Section \ref{sec:ideal} shows that
\begin{equation}
\deriv{}{t}{\flh(L_t)} = \Big[\phi + \avec\cdot\vec w - \psi \Big]_{x_0(t)}^{x_1(t)}.
\label{eqn:dadtN}
\end{equation}
Similarly, \citet{2018FlDyR..50a1408A} showed that \eqref{eqn:aly} becomes
\begin{equation}
\deriv{}{t}{\flh(L_t)} = \left[ U\pder{(\zeta + \psi)}{U} - (\zeta+\psi)\right]_{x_0(t)}^{x_1(t)}.
\end{equation}
Notice that $[\psi]_{x_0(t)}^{x_1(t)} = \int_{L_t}\vec E\cdot\dl$, which is precisely the parallel electric field used to define the reconnection rate in the theory of general magnetic reconnection \citep{1988JGR....93.5547S}. It represents the change in $\flh(L_t)$ due to a (non-local) change in the magnetic flux linked with $L_t$. The term $\avec\cdot\vec w$ is harder to interpret in general because it depends on the choices of $\avec$ and $\vec w$. With the gauge condition $\nhat\times\avec=\nhat\times\avec_P$ on the boundary, the $[\avec_P\cdot\vec w]_{x_0(t)}^{x_1(t)}$ term effectively represents a ``work done'' by motion of the field line $L_t$ with respect to the reference field \citep{2015PhPl...22c2106R}. 

For a given non-ideal evolution of $\vec B$, the field line velocity $\vec w$ is not uniquely defined, and consequently the identification of the field line $L_t$ over time is not unique. Whilst the component of $\vec w$ parallel to $\vec B$ is arbitrary -- as is clear from \eqref{eqn:ndecomp} -- it is the non-uniqueness of the perpendicular component $\vec w_\perp$ that changes the identification of $L_t$. To see that this component is non-unique, note that \eqref{eqn:ndecomp} implies
\begin{equation}
\vec w_\perp = \vec v_\perp = \frac{\vec B\times(\nabla\psi - \vec E)}{B^2}.
\label{eqn:wperp}
\end{equation}
On each field line we are free to specify an initial value of $\psi$. This does not change $[\psi]_{x_0(t)}^{x_1(t)}$ but does change $\vec w_\perp$.

One situation where a natural choice of $\vec w_\perp$ arises is the case when the field-line endpoints are line-tied on the boundary, meaning that $\vec u=\vec N=\vec E=\vec 0$ there \citep{2015PhPl...22c2106R}. In that case, we can identify $L_t$ over time by fixing one endpoint, say $\vec w_\perp(x_0)=\vec 0$ for all field lines. This is achieved by choosing $\psi=0$ throughout the region of the boundary where $\vec B\dotn<0$. At the opposite endpoints $x_1(t)$, the values of $\psi$ will then be fixed by $[\psi]_{x_0(t)}^{x_1(t)}$, leading to $\vec w_\perp(x_1)\neq\vec 0$ in general so that these endpoints will move in time in a non-ideal evolution. We could equally well fix $\psi=0$ on the region with $\vec B\dotn >0$, thus defining the field lines by fixed $x_1$ positions so that $x_0(t)$ varies over time. 

If $\vec u$ is non-zero on the boundary but $\vec N$ remains zero there (or can be neglected), then it is possible to subtract the ideal term $[\avec\cdot\vec u]_{x_0(t)}^{x_1(t)}$ from \eqref{eqn:dadtN} and isolate the change in field line helicity coming from non-ideal evolution. An analogous calculation was implemented for the magnetic winding measure by  \citet{2020gek} in data from a laboratory experiment of interacting magnetic flux ropes.

\subsection{Application to turbulent magnetic relaxation} \label{sec:relax}

The non-ideal evolution of $\flh$ has been explored only in the context of braided magnetic fields, where all field lines connect between two planar boundaries $z=0$ and $z=1$ at which $\vec u=\vec N=\vec 0$. The most significant finding to date is that, when the magnetic field line mapping from $z=0$ to $z=1$ is complex with sharp gradients, the evolution of $\flh$ for high $\mathrm{Rm}$ is dominated by redistribution between field lines, rather than dissipation \citep{2015PhPl...22c2106R,2021arXiv210801346Y}. 

\begin{figure}
\centering
\includegraphics[width=0.8\textwidth]{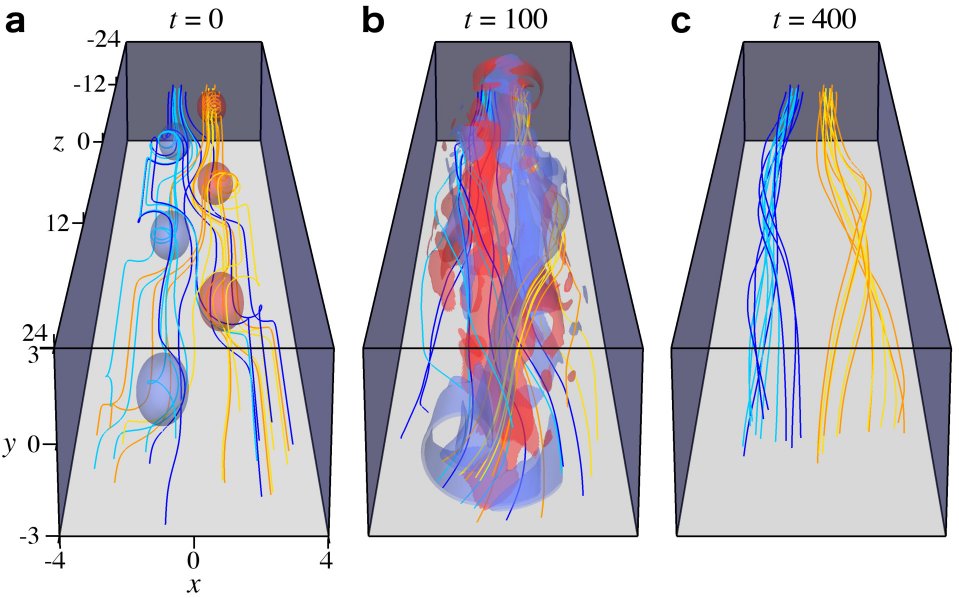}
\includegraphics[width=\textwidth]{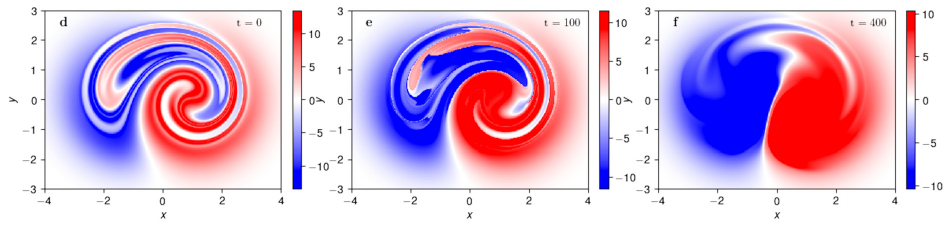}
\caption{Evolution of field line helicity during line-tied resistive relaxation of a braided magnetic field (see \citet{2021arXiv210801346Y} for details of the simulation). Panels a-c show magnetic field lines and isosurfaces of current density, in the initial condition (a), during the turbulent relaxation (b), and in the relaxed state (c). Panels d-f show cross sections of field line helicity on the $z=-24$ boundary  at the same times. Reproduced with permission from \citet{2021JFM...911A..30C}.}
\label{fig:braidsims}
\end{figure}

To see this, consider the evolution of $\flh$ on the field line traced from a fixed point $x_0$ on the lower boundary $z=0$. \citet{2015PhPl...22c2106R} made the natural choice $\vec w=\vec 0$ on $z=0$, in which case \eqref{eqn:dadtN} reduces to
\begin{equation}
\pder{\flh(x_0)}{t} =  \avec(x_1(t))\cdot\vec w(x_1(t)) - \psi(x_1(t)),
\label{eqn:dadt0}
\end{equation}
where $L_t$ is the field line defined by tracing from the fixed point $x_0$ at subsequent times. They argued that the $\avec\cdot\vec w$ dominates the $\psi$ term in a magnetic field with complex mapping, because $|\vec w|$ scales like $|\nabla\psi|/B$ from \eqref{eqn:wperp}. Since $\psi(x_1(t))$ is a field-line integrated quantity, and since the field line mapping has sharp gradients, we would expect $|\avec\cdot\vec w| \gg |\psi|$ for high $\mathrm{Rm}$. This was demonstrated originally in kinematic examples (see also \citet{mactaggart2019topics}), and has recently been confirmed by direct calculation of the terms in \eqref{eqn:dadt0} in resistive MHD simulations \citep{2021arXiv210801346Y}. One such simulation is illustrated in Fig. \ref{fig:braidsims}.

Watching the full evolution of $\mathcal{A}(x_0)$ in the Fig.~\ref{fig:braidsims} simulation suggests that the $\flh$ pattern evolves predominantly by rearrangement with some fictitious flow. To see this from the evolution equation, we can employ the seemingly unhelpful trick of labelling field lines by fixing their endpoint $x_1$ on $z=1$, so that $\psi=0$ and $\vec w_\perp=\vec 0$ on $z=1$. Then instead of \eqref{eqn:dadt0}, equation \eqref{eqn:dadt} becomes
\begin{equation}
\deriv{}{t}{\flh(L_t)} =  -\avec(x_0(t))\cdot\vec w(x_0(t)) + \psi(x_0(t)).
\label{eqn:dadt1}
\end{equation}
To compute the evolution of $\flh$ at a fixed position $x_0$, we must now account for the fact that the field lines are moving past this point with the local field line velocity $\vec w(x_0)$. Thus the time derivative in equation \eqref{eqn:dadt1} is a Lagrangian one and so at the fixed position $x_0$ we have
\begin{equation}
\pder{\flh(x_0)}{t} = -\vec w\cdot\nabla\flh(x_0) - \avec(x_0)\cdot\vec w(x_0) + \psi(x_0).
\label{eqn:dadt1w}
\end{equation}
Notice that all of the terms here are ``local'', although $\psi$ and $\flh$ are still field-line integrated quantities. The advection term $\vec w\cdot\nabla\flh$ is effectively a product of two gradients of field-line integrated quantities, so one might expect this to be largest. It is therefore no surprise that the dominant behaviour observed is rearrangement of the $\flh$ pattern by advection.

\citet{2021JFM...911A..30C} took this idea and developed a variational model for turbulent magnetic relaxation that predicts the relaxed state to have the ``simplest'' pattern of field line helicity achievable by pure advection. This already predicts the two oppositely-twisted magnetic ``flux tubes'' seen in Fig.~\ref{fig:braidsims}(f). However, closer inspection of the numerical simulations by \citet{2021arXiv210801346Y} shows that the other terms in \eqref{eqn:dadt1w} do play a role in establishing the substructure of the final state. 

\bibliography{helicity}

\printindex

\end{document}